# Evolution with hole doping of the electronic excitation spectrum in the cuprate superconductors

**John A. Wilson**

H. H. Wills Physics Laboratory
University of Bristol
Tyndall Avenue
Bristol BS8 1TL.  U.K.

**Abstract**

The recent scanning tunnelling results of Alldredge *et al* on Bi-2212 and of Hanaguri *et al* on Na-CCOC are examined from the perspective of the BCS/BEC boson-fermion resonant crossover model for the mixed-valent HTSC cuprates.  The model specifies the two energy scales controlling the development of HTSC behaviour and the dichotomy often now alluded to between nodal and antinodal phenomena in the HTSC cuprates.  Indication is extracted from the data as to how the choice of the particular HTSC system sees these two basic energy scales ($U$, the local pair binding energy and, $\Delta_{sc}$, the nodal BCS-like gap parameter) evolve with doping and change in degree of metallization of the structurally and electronically perturbed mixed-valent environment.

**§1.  Introduction regarding appraisal of the ARPES and STM results for HTSC cuprates.**

A steady stream of technically advanced ARPES and STM based papers concerning the cuprate high $T_c$ superconductors are in the view of this author presently being seriously misconstrued.  While the work is intrinsically of great value to the understanding of the HTSC problem, the proffered treatments of this data would appear continually to compound certain initial misconceptions.  These include that, in both experimental areas, the peaks dominating the spectral onsets represent 'coherence peaks' such as are to be found in similar work on classical superconductors.   I have endeavoured in earlier papers [1] to signal that this cannot be the case for the HTSC materials.  The perspective from which the latter appraisal has been reached is one of boson-fermion resonance within an inhomogeneous negative-$U$ circumstance, as is deemed appropriate to the mixed-valent cuprate systems.  The essential mixed valence is introduced either by cation and anion substitution or by the insertion of excess interstitial oxygen ions.  The electronic effects of these chemical/charge alterations are here highly local in consequence of the host divalent compounds being uniformly Mott insulators.   Nothing actually better highlights the prevailing structural and electronic inhomogeneity than do STM results.  From such work it is evident spatial conditions vary very significantly, virtually at the unit-cell level [2].  These very fine scale structural and electronic inhomogeneities 'active' within HTSC materials have long been in evidence, below as above $T_c$, from the probings of EXAFS, $\mu$SR and NMR.  However some researchers persist in regarding the exposed inhomogeneity as being an extrinsic or surface-based matter, in view of the fact that the electrical, magnetic and thermodynamic data relating to the superconducting state all appear to support a very 'clean' situation as existing within carefully prepared samples.  $T_c$ is indeed sharply defined, and optimal $T_c$'s are consistently attained as the hole doping is brought through to $p$ = 0.16 [3], an observation early on formulated by Tallon *et al* with their quadratic expression for $T_c(p)$ — in BSCCO-2212  $T_c(p)$ = 95K × {1 – 82.6$(p – 0.16)^2$}.  That the onset of superconductivity is globally so well-defined at a time when the quasiparticle behaviour stands deeply perturbed is one of the most striking aspects of HTSC behaviour.  It slowly has become accepted that the Fermi surface in the HTSC materials ($T > T_c$) suffers an advancing state of disintegration as one travels out onto the axial arms of the computed surface [4].  This disintegration has been much discussed in terms of 'hot-spot' scattering, centred upon where the FS is intersected by the oblique diagonal within each Brillouin zone quadrant (see fig 1).  The scattering has been widely couched in terms of spin fluctuation scattering, often in association with incipient SDW or dDW formation.  I have, by contrast, repeatedly expressed that I perceive this key scattering to involve local pair formation and annihilation in a resonant, boson-fermion crossover circumstance [5].  The condition above $T_c$ then has to be regarded as an integral part of the HTSC phase change.  This integrality is more complex however than one simply of local pairs pre-existing above $T_c$ which proceed below $T_c$ to acquire mutual phase coherence.  Many experimental results, such as those from EELS [6], neutron scattering [7] and Nernst [8] work, clearly support some form of extrapolation of superconducting-like behaviour through to considerably above $T_c$. In the current scenario of resonant negative-$U$ crossover, the change though which $T_c$ demarcates is not one simply between local pairs and their own low temperature bosonic condensation, but rather one between pre-existing local pairs and a condensation of the entire electronic complement around the full Fermi surface. Here the condensate is a global one, with $T_c(p)$ fixed and the superconducting state established over both the local-pair negative-$U$ bosons and the induced Cooper pairs alike [9]. The symmetry of the driven, 'inner', BCS gapping is well-established (*eg.* from Tsuei *et al*'s fractional vortex observations [10]) as being $B_{1g}$ $d_{x2-y2}$- in geometrical form, but it is my understanding that the symmetry associated with the 'outer', local-pair, gapping takes extended-*s* or $A_{1g}$ form, similar to the square of the former [1b]. The induced condensate in this way determines phenomena around the nodal 45° directions, whilst the local-pair (LP) physics dominates on the axial arms/saddles of the computed Fermi surface, in particular near the 'hot spots'. A point to be registered here, and one followed up on below, is that LP stability relative to $E_F$ does not in fact extend through to the 45° nodes. Indeed a strong 'pseudogap' LP signal is tracked in the energy-resolved STM work only roughly between the hot spot locations and points 15° from the neighbouring node [1a,4] (the angle measured being here about the B.Z. corners around the near-circular F.S.).

Most significant in the above is that while the superconducting state itself runs extensively (in phase if not local magnitude) throughout a quality sample, the pseudogap physics, which derives from local pair formation and annihilation, is much more inhomogeneously structured particularly above $T_c$. This latter inhomogeneity is dictated by the frozen dopant atom array. Usually the array of substitutional or interstitial ion sites in these systems is close to being random, but, as is now well documented, there exists some tendency (largely under the constrictions of the Jahn-Teller effect) for the associated electron holes to order dynamically into charge/magnetic 'stripe' arrays [11,5d]. Such fluctuational organization of the carriers is not in fact registered in the STM work, which presents a 'static' (rastered and binned) image fixed by the dopant ion distribution itself. That does not however gainsay appreciable electronic self-organization from occurring dynamically and from being of relevance to the detailing of the final state acquired – the globally supported superconductivity. It simply means that the intensely local STM probe is not suited to registering any fluctuating spatial order. This requires the *k*-space, correlation sensitive, probing of electron and neutron *diffraction*. Care must be taken at this point to distinguish between where there is a Fourier transform aspect to the primary probe and where simply a Fourier transform is afterwards secured by optical means from the static primary probe map in order to gain periodic information not immediately evident within the general chaos of the map. The latter procedure has successfully been followed with the STM tunnelling current maps to reveal (very easily) the incommensurate superlattice period in BSCCO-2212, and with more difficulty there the checker-boarding distortions (so strong in more ionic Na-CCOC) [11a]. Of greater interest to the present paper are the sets of still more diffuse static periodic imprint existing in the frozen 'gap maps' of the energy-resolved STM work (see §2 below) and that directly impact on the superconducting condition.

The above strong collisional activity between fermions and fermions and between fermions and bosons is viewed by the author as being behind all the highly characteristic electronic properties of HTSC materials [12,5]. The fermion-fermion (to boson) hot-point scattering leads to a $T^2$ resistivity contribution, and dominates at high doping where local pair stabilization has become less entrenched, due to reduction in the binding and definition of the negative-$U$ centres under the increased screening there. By contrast, back where pair formation is optimized around $p \sim$ 0.16-0.18, the observed T-linear resistance conveys the predominance here of fermion-boson scattering. What particularly is to be emphasized is not just the functional forms to the above electronic scattering but the high intensity and wide range of temperature applicability. The observed steady transfer of the scattering between $p$ = 0.15 and 0.25 from being dominantly $T$-linear, quasi-elastic and anisotropic for $T_c^{opt}$ material, to being quadratic, inelastic and isotropic as $T_c$ recedes toward the OD limit entirely matches one's expectations for the local pair model and a resonant crossover with $E_F$.

As long has been apparent for the HTSC cuprate systems, and now is again so clearly in evidence from the ARPES data beyond $T_c$, it is the above intense scattering which is responsible additionally for the abnormal 'normal state' temperature dependence both of the Hall [13] and Seebeck [14] coefficients. The drive experienced within HTSC cuprates towards weak localization and to a surrender of quasiparticle coherence consistently expresses their level of proximity to the parent, undoped, state and Mott insulation. The loss of coherence is accordingly observed to expand rapidly upon sub-optimal hole underdoping, this the more so the more ionic the HTSC system involved. In optical work of various types it is discovered moreover that the chronic scattering climbs with the excitation energy/photon frequency once more in closely linear fashion near optimal doping [15]. The above strong scattering falls away rather rapidly below $T_c$, as long evident from NMR relaxational rate studies, but it remains significantly high even below $T_c$ until the majority of quasiparticles have become taken into the superconducting condensate. Notable similarly in our current mixed situation of instigating local pairs and induced Cooper pairs is that a substantial proportion of the local pairs remain outside the general condensate to well below $T_c$. These excited local pairs bear non-zero crystal momentum and find themselves incorporated into a mode of energy versus pair momentum rising linearly from just above where the ground-state binding energy maximum [16] is located (on the former F.S. arms near the hot spots, defined as previously - see fig.3 in [5b]). Such non-condensed local pair states have been ascribed by the author [5c] to be source of the extra conductance uncovered by Corson *et al* within the microwave part of the spectrum [17]. This particular conductance is additional to that of the broadened delta-function microwave response from the superconducting condensate itself, and again too to that of the residual Drude behaviour from the uncondensed quasi-particles, in evidence in far I.R. work [18].

Because of the above complex, inhomogeneous and highly fluctuational nature of the superconducting state, it would be highly surprising then for the superconducting condensate in standard fashion to generate sharp, undamped, coherence peak containing, energy spectra. After all, the condensate gapping is at its maximum in the near-axial parts of the FS, well away from the 'quiet' nodal regions of the anisotropic superconducting state. Indeed the gap maxima occur precisely where, above $T_c$, the endemic strong scattering succeeds in eliminating the very integrity of the Fermi surface [4]. It would be most odd accordingly for the coherence of the antinodal (AN) superconducting condition to stand less perturbed than occurs closer in to the nodal (N) direction. Yet that is what both the ARPES and STM data would imply if, as customarily is assumed, the recorded spectral peaks are taken simply at face value to be standard coherence peak structures. I have indicated previously in [1] that this cannot be the correct interpretation.

A prime fact at this point to appreciate is that even near optimal doping the peakings recorded by the ARPES and STM techniques behave neither in classical fashion, nor indeed in quite like manner to each other [1b,19]. The gap values implicated are already appreciably greater than match the BCS expression for d-wave mean-field behaviour, namely $2\Delta_{sc}(0) = 4.3kT_c$, to which other modes of assessment of $\Delta_{sc}(0)$ (like penetration depth, low temperature specific heat and $B_{2g}$ Raman) more closely adhere. The above discrepancy rapidly becomes ever more marked into the underdoped region, where as $T_c$ drops away the seemingly coupled maximal gap value in evidence grows so rapidly as to render it self-evident that this assessed gap and its associated peak feature are not conventionally related to $T_c$. Many techniques actually have registered DOS gapping activity above $T_c$. All such action has tended to become lumped under the broad umbrella of 'pseudo-gap behaviour'. This extended gapping potentially can accrue from many quarters beyond that of prime interest here – local pair formation. One may well foresee contributions arising from a susceptibility to Mott insulation, to local magnetic moment formation, to stronger spin fluctuations, to incipient standard SDW and CDW formation under Fermi surface nesting, and finally to charge ordering into stripes, etc. Such striping in fact continues over into the corresponding Mott-insulating manganates and nickelates, and is largely governed by Jahn-Teller effects within the *d*-electron states and relating to local structural bonding. Such structure and bonding changes contribute strongly likewise to charge disproportionation phenomena on display in a compound like $AgNiO_2$. The effects sometimes are sufficiently great to transform an otherwise metallic compound such as 5*d* AuO or $CsAuCl_3$ through into becoming a band insulator [5d].

If the dominant peaks in the superconducting state spectra observed in the ARPES and STM work are not to be understood as conventional coherence peaks, how are they to be regarded? Clearly both features manifest direct connection with the superconductivity. In [1] I advanced the following resolution of the matter.

The pseudogap observed here (initially with and ultimately without attendant peak) grows monotonically in magnitude with underdoping, as witnessed in the fairly recent ARPES papers from Tanaka et al [4c] and Valla et al [20]. This pseudogap energy I have taken in [1] to monitor the steady growth in binding energy (cursive $U$) of the negative-$U$ pair state below $E_F$, under the diminishing metallicity. There is less screening out now of any local trivalent Madelung site potential wherein the local pair states are created. However, as into the underdoped regime the binding energy $U$ of the local pairs grows, so simultaneously will mount and diversify such site relaxations and their specific energy. Given this, not to mention the diminishing pair population, and the state's spectral definition progressively degrades.

In this way one can understand how the witnessed and somewhat divergent behaviour comes about for the so-called 'coherence peaks' seen in the high energy photoemission and the low energy scanning tunnelling experiments. The form and sourcing of these features differs in detail, and in neither case do the experiments make measure of a superconducting gap as present in a traditional BCS material. The spectra stand dominated in each case by peaks that to varying degrees fall beyond this energy. As (2)$U$ and (2)$\Delta_{sc}$ draw together upon the approach to local pair resonance at or rather just below $E_F$ at optimal doping, it becomes easy mistakenly to conflate these two aspects to the resonant global superconducting state, and to treat the peaks detected as standard coherence peaks.

I do not wish at this point to expand further upon this but to turn to examine closely the experimental results contained in two very recent papers. That should illustrate greatly what is being stated above. The new works are (i) the location-resolved STM paper on BSCCO-2212 from Alldredge *et al* [21] covering a wide range of average sample dopings <$p$>, and (ii) the Fourier-transformed energy-resolved STM paper on Na-CCOC from Hanaguri *et al* [22] (for comparison with earlier such work on BSCCO-2212 from McElroy *et al* [23] and Hoffman *et al* [24]). Neither new paper is especially easy to read each being rather condensed and, I would claim, regrettably misrepresented in the manner pointed to above.

**§2. A close examination of the new STM results from Alldredge *et al* on BSCCO-2212.**

The STM paper from Alldredge *et al* [21] presents angle-integrated, energy-resolved, tunnelling current spectra from Bi-2212 samples of six $T_c$-assessed compositions, $p$ = 0.08, 0.10, 0.14, 0.17, 0.19 and 0.22 (*i.e.* appreciably underdoped to appreciably overdoped), the associated $T_c$ values running from 65 K, over optimal, and back to 65 K. From these a scan set of more than a million readings of the local tunnelling conductance $g(r,V)$ has been taken, rastering across a spatial field of operation of 40 nm x 40 nm and at tunnelling voltages running between 0 and ± 100 mV. This great size of data bank is called for by the fact that the tunnelling conductance alters so markedly across the face of any given sample, requiring spatial sampling density effectively at the unit cell level [2]. The $g(V)$ conductance signals are collected every 2 mV. Individual $g(V)$ spectra from set positions $r'$ constitute the primary data plots – note the individual data points appear on *supplementary* figures 2b to e in [21] for the $p$ = 0.10 sample at five different locations $r'$. The enormous variability here from spectra with sharp peaks beyond the central dip, through to no peaks, and even no specific edge feature is immediately evident. With 64,000 individual such traces the immediate problem is how to convey the evolution in global behaviour that occurs with change of sample doping composition $<p>$. The manner in which this now is being done by Alldredge and coworkers is however, to the present author, introducing real conceptual problems.

By focussing upon the clearest feature in these spectra, the peak/edge, Alldredge *et al* have been drawn (in line with the group's earlier work) to fit the data to the BCS superconducting procedure of their equation (1) for $N(E)$, plus an empiric extension now to equation (2). The approach stands in compliance with the traditional outcome which they display in figure 1. This figure, with its BCS coherence peaks and progressive damping, elaborated now for *d*-wave symmetry, constitutes the conventional picture to which the various data sets are here made to conform to grant extraction of 'hard numbers' relating to gap size, damping and ±V tunnelling asymmetry. Note, however, in this introductory figure the limiting (peak-to-peak) superconductive gap value $2\Delta_{sc}$ at 4.2 K is portrayed holding 'favoured' magnitude 40 meV, whereas in the actual data, even from the slightly over-doped $<p>$ = 0.17 sample, rarely do the empirical peak-to-peak gaps drop close to such a value – see the gap map $\Delta(r)$ of fig.3c in [21], where red or even orange $\Delta$ regions scarcely ever appear. Already at this doping the sample is becoming dominated by whole regions for which the local gap maximum is shifting to the considerably higher values of the pseudogap regime. Furthermore note by $<p>$ = 0.14 (fig.3b) how the gap map has become mainly green or even white (where $\Delta$ stands higher than for green but no longer with any peaks to measure to), meaning $\Delta_{pp}^{(exp)}$ (and implicitly here $\Delta_{sc}$) is now almost everywhere larger than 80 meV. Clearly the pseudogap has become predominant.

The general situation becomes rendered still more problematic as one imposes equation (2) to procure *fitted* gap values $\Delta_1^{(fit)}$. Notice then how fig.3f bears more yellow than its empiric partner at $<p>$ = 0.22 (fig.3a), fig.3g more yellow than fig.3b for $<p>$ = 0.19, and fig.3h more green and blue than fig.3c for $<p>$ = 0.17. This numerical massaging of the data is masking its true message, and the process becomes yet further compounded in figure 2 where fit-derived spectra, from averaging over all those $r'$ to have yielded common fit value $\Delta_1^{fit}$, are set out as the leading data of the paper, whilst the true primary material is hidden away in the Supplement. In figure 2 in [21] the various spectra across the $<p>$ = 0.10 sample, after being averaged together according to their $\Delta_1^{fit}$ values, are presented overlaid by their grand best-fit lines. Over and above the data smoothing which has occurred here, we notice that by concentrating upon the peaks ($2\Delta_1$ takes here, note, the *minimum* value of 76 meV) we have not in fact arrived at a particularly good fit at smaller energies, in particular for those data sets for which $2\Delta_1$ is least. If one examines fig.S3, where 17 such averaged spectra from four samples ($<p>$ = 0.22, 0.19, 0.17 and 0.10) have been arrayed sequentially by $\Delta_1^{fit}$, we can observe directly the dangers inherent in the treatment employed and in the further data handling to come in fig.4 onwards. In fig.S3 the circle sequences on traces 4 to 8 begin to look *s*-wave in form and quite unlike the globally imposed *d*-wave fits shown. Only for trace 9 and beyond, in which $2\Delta_1^{fit}$ stands in excess of 100 meV, does a single *d*-wave form look to provide appropriate fitting.

The primary data clearly demand to be treated, and indeed to be regarded, very differently. There arrives in due course within the above paper a slow acknowledgement that the data in fact hold more than was first presumed in the analysis, and that this involves the existence of a second feature, labelled there $\Delta_o$. However, as Alldredge *et al* subsequently handle this weak inner feature via data dealt with in the above outlined fashion, its revelation is blurred. When one returns however to the primary data of fig.S2 (probably for <*p*> = 0.10) one finds this feature, which lies close to 18 meV, stands quite well-defined in spectra for which $\Delta_1^{fit}$ is least, it fading steadily away wherever $\Delta_1^{fit}$ passes beyond 50 meV. This $\Delta_o$ now exhibits the correct characteristics to take on the true gap designation $\Delta_{sc}$ appertaining to $T_c$ – in the view of the present author the induced superconductivity secured by the Feshbach resonance of the antinodally formed local pairs. Over *all* locations *r'* this empiric inner edge possesses effectively the *same* energy and it presents, moreover, no strong coherence peak. It is the outer, much stronger feature that manifests intense positional variability, appropriate to the presence of a local pair-binding potential $U(r)$ established in the random distribution of mixed-valent centres (as portrayed in figure 4 of [25]). Recall that the binding energy $U$ of the local pair state below $E_F$ is but the final small fraction of a very much greater negative-$U$ energy to have effected translation downward so strongly of the double-loading state $(p^6)d^{10}$ that comes with complete energy relocation at shell closure of the hitherto bonding and antibonding ($pd\sigma/\sigma*$) states [5b fig.1 and 25 fig.3].

The key empirical observation of the Alldredge paper is then of a weakly damped, homogeneous, inner gap existing in conjunction with a heterogeneous outer gap. And the outer gap is the one specifically associated with the intense scattering behaviour; scattering which is linearly dependent upon $E$ and markedly asymmetric in ±$V$, and each feature developing rapidly as *p* is taken below optimal doping.

The above work by no means is the first occasion on which low temperature tunnelling experiments have given evidence of a compound gap structure appropriate to upholding an altogether more intricate HTSC scenario than is all too frequently being adopted. The standard *c*-axis tunnelling work of Krasnov, Ozyuzer, Zasadzinski and their coworkers [26] provided clear evidence of this phenomenon and of the changes encountered with temperature and doping. The early STM work of Maggio-Aprile, Fischer *et al* [27] by probing inside magnetic vortices tellingly separated the two gapping processes; only the outer gapping feature outlives the loss of superconductivity within the vortex. Moreover this intimate two-stage density-of-states gapping has long been in evidence thermodynamically from the specific heat work of Loram *et al* [28]. The entropy analysis which they made allows the progressive development of these two gaps to be followed as functions of temperature and doping. In underdoped HTSC material it is evident that the pseudogapping is a much more robust feature than is the final superconductivity engaged with at $T_c$. The stabilization arriving below $T_c$ in free energy is found for every HTSC system examined [28] to diminish sharply below a *p* value of around 0.18 – a result I have commented upon at length in [11a,b]. Oda *et al* [29] have pursued such studies in the light of more recent knowledge that with underdoping the *BCS-like* component to the overall superconducting condition below $T_c$ becomes more and more confined to the nodal regions. This arises in register with the growing break-up in coherence suffered by the Fermi surface in its outer reaches under the chronic scattering in evidence there both in transport and angle-resolved photoemission data. Kanigel *et al* [30,4a] a little while ago provided a very illuminating ARPES study of this dichotomy in behaviour of the nodal and antinodal signal across $T_c$. While gapping on the inner third of the F.S. near the nodal position is seen to collapse at $T_c$, the outer two thirds around the hot spots continue to experience DOS diminution almost to room temperature. This impairment to the outer segments of the F.S., and hence to 'normal' state coherence, becomes much more marked with increase in the ionic character of the system investigated, either from reduction in hole doping or by increase in counter-ion ionicity, as is most apparent when working with $Ca_{2-x}Na_xCuO_2Cl_2$ (Na-CCOC) [31]. While two-gap behaviour in YBCO and Bi-2212 is not immediately evident for near-optimally doped material because the two energies have moved together, for all systems where $T_c$ is less successfully effected this confusion quickly is removed. At the ionic limit we have just pointed above to Na-CCOC, where $T_c^{opt} \sim 25$ K, and a complementary situation occurs with covalent Bi-2201, for which $T_c^{opt}$ again is down around 25 K. Kondo *et al*'s and Boyer *et al*'s ARPES work on the latter material [32] and Tanaka *et al*'s on strongly underdoped Bi-2212 [4c] each supply very clear evidence of the considerable separation now in the energies relating to nodal and antinodal gapping. We will return in §3 to examine how these relative changes develop systematically across the full range of HTSC materials.

All this experimental information has not, it is felt, been adequately embraced in the STM paper from Alldredge *et al*. I previously have taken issue in [1a] with the analysis of the STM results proffered in [23] and [24] over the conflation of the two aspects to the HTSC event. In particular the problem revolves around repeatedly trying to present the large $\Delta$ energies extracted as relating directly to $T_c$, when clearly they amount often to considerably more than matches $5kT_c$. It is a confusion that has become entrenched in the field largely through the ARPES results, and which is still on show in the key ARPES work of Kanigel *et al* on underdoped BSCCO-2212 [3]. It is better probably to refer to the HTSC circumstance as a two-stage or two-component process rather than a two-gap process, where the latter might refer to conflicting states rather than to two aspects of the one resonant process. Most certainly, of course, it can in no way be maintained as a one-gap process with simple BCS-like character. Such misconception becomes especially unfortunate in the Fourier transform STM work pioneered on BSCCO by Hoffman, McElroy *et al* [23,24] and extended recently to Na-CCOC in a new paper from Hanaguri *et al* [22]. The latter work we shall now turn to examine in detail in §3. It is the current author's claim, as made previously in [1a], that one most certainly here is not looking at what originally was suggested in [23,24] – standard Bogoliubovon dispersion.

**§3. A close examination of the new STM results from Hanaguri *et al* on Na-CCOC.**

The type of experiment undertaken is more advanced in concept than that above from Alldredge *et al* [21]. Hanaguri *et al*'s paper [22] follows the earlier procedure of Hoffman *et al* [24] and McElroy *et al* [23], and involves picking up spatial periodicities in the STM topographic image by using optical Fourier transform investigation of that image. The previous work was performed on near-optimally-doped Bi-2212, but attention in the new paper has been transferred to Na-CCOC ($Ca_{2-x}Na_xCuO_2Cl_2$) at the opposite end of the ionicity scale for HTSC systems. The $T_c$ value in the towards-optimal samples of Na-CCOC now employed ($p \approx 0.13$) is down near 25 K, and the system presents much stronger pseudogap behaviour than BSCCO. As well as the gapping relating to superconductivity and local pairs, one in Na-CCOC encounters further gapping due to strong checker-boarding, with DOS loss extending now out to magnitudes beyond 200 meV. I have already discussed in [11a] how I believe checker-boarding to ensue from F.S. 'boxing and nesting' in a highly correlated form of CDW behaviour, the outcome of physics expounded in [33] by Khodel and others. In BSCCO by contrast, checker-boarding is a weak phenomenon even for rather underdoped material. The dominant real space periodicity actually apparent to Fourier transform STM investigation of BSCCO-2212 is the incommensurate misfit structure [34], of approximately $5b_o$ along the orthorhombic ($\sqrt{2}a_o$) axis in the 45° direction of a pre-distorted single $CuO_2$ layer. Na-CCOC (related to LSCO) is a much more rigid system and it exhibits no such superstructure. What has additionally been detected in BSCCO is a third weaker source of spatial periodic disturbance, and this in the new work has now been comparably revealed for Na-CCOC. This time the lattice modulation bears evident and highly revelatory connection to the superconductivity.

We will first introduce discussion concerning this further source of periodic structural disturbance exposed in the optical Fourier transform by reference to Hoffman *et al*'s [23] and McElroy *et al*'s [24] pioneering work on BSCCO, before proceeding to Na-CCOC and the new work of Hanaguri *et al* [22]. This is because with the latter material the extraction of the results becomes somewhat more intricate in consequence of the simultaneous presence there of the strong checker-boarding.

It was noted that the wavevectors of the first of the new disturbances detected, labelled $q_1$ and $q_2$ in [23], spanned respectively across and between the arms of the Fermi surface to symmetry equivalent positions. As the *p* value of the sample was increased (*i.e.* its electron count was decreased) it was observed that $|q_1|$ is accordingly decreased and the associated $|q_2|$ correspondingly increased. Recall these tunnelling experiments were performed at He temperatures, deep in the superconducting regime, and for any one run they are made at select voltage *V'*. It soon was recognized that as the tunnelling voltage progressively was changed the locus of ($q_1$,$q_2$,V') was monitoring around the Fermi surface the magnitude of the superconducting gap there, as if emulating what hitherto had been directly recorded only by ARPES. It was argued by the authors of [23] and their associates that what was being sensed by STM were the Bogoliubov dispersion curves as they would appertain to a standard superconductor – though in this case of *d*-wave symmetry.

The original detection of the above phenomenon by Hoffman et al [23] related just to the two modulation vectors $q_1$ and $q_2$, but duly it was recorded by McElroy *et al* [24] that all seven wavevectors joining (certain) $k'_F$ to symmetry equivalent locations on the FS were 'active' for each chosen tunnelling voltage $V'$. These disturbances were accordingly to be viewed as standing waves, set up under coherent elastic scattering and possessing these interconnecting wavevectors, a process referred to in [23,24] as QPI or quasi-particle interference. The scattering directly off the Bogoliubovons, and specifically reflecting $\Delta(\mathbf{Q}, V')$, was to be linked with the FS geometry in compliance with the density of states – the 'bananas model'. Figure 2a indicates how for the given geometry the seven elastic scattering vectors $\mathbf{q}_1....\mathbf{q}_7$ from a particular starting wavevector $\mathbf{k}_F'$ are disposed. Each of the 8 such equivalent starting locations generates likewise its own circuit of 7 further such vectors, thereby establishing (8x7)/2 independent links in all. These 28 $\pm\mathbf{q}$ scattering linkages see created 28 independent standing waves, and those stationary waves then proceed to source the 28 diffuse satellite diffraction spots in evidence in the optically obtained Fourier transform of the topographic STM map $g(\mathbf{r}, V')$ (fig.2b).

However, as I pointed out earlier in [1a], there is something seriously wrong with the above interpretation: (*i*) it neglects the fact that only a limited fraction of the near-circular Fermi surface is traced out within any one BZ quadrant (in BSCCO that between $\phi \sim 15°$ and $30°$); (*ii*) the as-extracted dispersion of the alleged quasiparticle gapping is much closer to being linear than it is to following the sought after *d*-wave form; (*iii*) the energy distribution of the sensed features stands (as with the ARPES data to which it attempts to relate) considerably in excess of $2\Delta_{sc}(\phi = 0°)$ being commensurate with. 5½ $kT_c$ or 20 cm$^{-1}$. Indeed even around the hot spot the assayed value of $\Delta$ exceeds 30 meV. This, in keeping with §§1 and 2 above, implies that it is a measure not of BCS-like gapping but of $U$ the local pair binding energy.

Now in [1a] I have advanced that what is being sensed in this STM experiment are not the Bogoliubovons associated with a BCS-like, single-component circumstance, but the uncondensed local-pair modes of a two-component circumstance associated with negative-$U$ driven BCS-BEC crossover and fermion-boson coexistence. According to the B-F crossover analysis performed by de Llano and coworkers [35] the dispersion of the uncondensed bosons is linear and the mode rises from the *K*-space region of maximal local-pair formation energy. In fig.2 of [1a] I indicated how to present this latter mode as accommodating the STM-derived data, extending from close to the hot spot angle of $\phi \approx 15°$ up to near $\phi \approx 30°$ at which point the linear mode crosses $E_F$. From the Fourier transform results no indication is given of the dispersion actually pulling around to an $E = 0$ gap node at $\phi = 45°$.

One key feature to have drawn Davis and coworkers [22-24] to contemplate the Bogoliubovon scenario was the observed particle-hole symmetry of the STM results. Apart from a to-be-expected difference in the spectral broadening, the Fourier transforms extracted from the $g(\mathbf{r}, V)$ maps are effectively identical for tunnelling voltages $\pm V'$. This brings to mind the ARPES spectra [36] where one gains the impression of quasiparticle band back-bending symmetrically disposed across $E_F$ (for each and every embraced $\mathbf{k}_F'$) within the superconductively gapped state, for the participatory electron and hole states alike – the classic Bogoliubov construction for the standard BCS circumstance. However de Llano *et al* [35] have pointed to a neglected particle-hole equivalence that they claim is being entirely suppressed within the conventional treatment of superconductivity. Namely that along with electron pairing, 2*e*, there always occurs concomitant hole pairing, 2*h*. This is more general than the Bogoliubov treatment of the BCS limit, and ought to apply to the crossover situation for single fermions and local pairs under the negative-*U* engendered resonance deemed as operative here in the mixed-valent cuprates.

It is an interesting question as to whether in these circumstances the tunnelling is more likely to involve pairs directly or whether the pairs subsequently associate. The STM experiments could with advantage be repeated using a tip of some appropriate superconducting metal. Already when employing *c*-axis junctions in standard tunnelling a duality of outcome has been observed [37,26a]. This question has resurfaced recently in one of the very few theoretical works drawn to probe into such matters [38]. This paper from Tornow *et al* addresses the more restricted problem of non-equilibrium two-electron transfer within a model redox system covered by a two-site extended Hubbard model embedded into a dissipative environment – the sort of model that might describe the incipient disproportionation Pd(II)/Pd(IV) in a system like PdTeI. In the donor/acceptor terminology employed in that paper, of the four different conditions $D^{2-}A^0$ (doubly-occupied donor), $D^0A^{2-}$ (doubly-occupied acceptor) and $D^-A^-$ (two singly-occupied, doubly-degenerate states), in respect to our crossover process [1,5,25] if we make the formal equivalence regarding two-electron transfer one may appreciate the relevance of their analysis. Beyond our above encounter with spatial periodic disturbance, one finds now in Tornow *et al*'s work the possibility also of temporal ringing between the pair states in the non-equilibrium dynamics. This obviously is a line to be probed in the future, possibly in connection with 1/*f* noise studies [39].

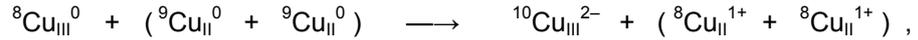

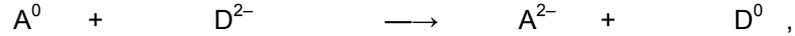

Let us now return to the details of Hanaguri *et al*'s work on Na-CCOC [22]. In order to eliminate the effects of checkerboarding upon the g($\mathbf{r}$,*V*) map, direct use has been made of the above relationship between the +*V* and –*V* topographic images and their FTs. By appealing to a likely phase difference between outgoing and scattered waves above and below $E_F$, instead of analyzing g($\mathbf{r}$,*V*) directly, analysis is pursued of the conductance *ratio* map g($\mathbf{r}$,+*V*)/g($\mathbf{r}$, –*V*) ≡ Z($\mathbf{r}$,*V*), and of its FT |Z($\mathbf{q}$,*V*)|. The 'residual' FTs so secured are indeed then found to display very similar characteristics to those obtained more directly from Bi-2212. However the details now make it very apparent that something is amiss with the Bogoliubovon based interpretation being offered. The results provide very illuminating insight into how in fact choice of the HTSC system for investigation brings significant modification to the detailed observations.

In BSCCO-2212 the data points extracted relate to the angle $\phi$ falling between 15° and 30°. Now in Na-CCOC the data points run from $\phi$ ~ 23° to 35°, much further from the hot spot and over towards the 45° node. Clearly in [22] there arises temptation to skew these data points round towards the node, but even then what is indicated in their fig.4 is not of classic *d*-wave form. One can on the other hand contemplate a 2-component decomposition, with a very short BCS nodal segment associated with a 5½ $kT_c$ equivalent gap magnitude of $2\Delta_{sc}$ = 12 meV (*ie.* a $\Delta_{sc}$ of just 6 meV). The rest of the data points clearly are associated with a much steeper dispersion and larger gap value. The dispersion shown by the bulk of the data points is in fact now appreciably steeper than found in the BSCCO-2212 case. If we extrapolate the effectively linearly dispersed data points back across the now large distance to the zone boundary we arrive at an energy of 65 meV or more – at least 10 meV greater binding than for optimally doped BSCCO-2212 and from a material now with $T_c$ only one third the size.

In figure 3 this comparison between the two materials is displayed, along with, on the same scale, the projected behaviour for a more covalent system such as Bi-2201. As the covalency (*ie.* screening) rises, so the binding energy of the local-pair zone-boundary state diminishes and the mode dispersion becomes less steep. Indicated as well in the panels is *d*-wave BCS-like gapping corresponding to $\Delta_{sc}(\phi)$ for the optimal circumstance $2\Delta_{sc}(\phi = 0) = 5½ kT_c$. In Bi-2212 one encounters crossover between the two states mid-way round the FS quadrant, but in the other two systems this crossover occurs closer to the node. This expresses in fact that boson-fermion coupling will be less extensive and effective at engaging all the quasi-particles and at raising $T_c$ in systems for which $U$, the pair binding energy, is either too small or too great to uphold optimal resonance.

In figure 4 a schematic model has been developed of how the fanning out of the local-pair mode states steadily augments in each system, upon increase in metallic screening, as the operative *p*-value goes up. In Na-COCC, with the smallest degree of fanning, $U(p)$ stays always far below $\Delta_{sc}(p)$, whereas in "Bi-2201", for $p$ = 0.28, $U(0°)$ has passed above $E_F$, and the pair state then is unbound. The $U(\phi = 0°)$ value to which, under this schematic, it would appear to tend, within all three panels, as $p$ is taken to zero, is ≈ 85 meV. Gaps greater than this are to be attributed to other physics like checker-boarding, striping, or Mott insulation. Note this energy is the same as that of the 'waterfall effect' discovered in recent ARPES results [40] on underdoped material, where clearly ultra-strong correlation effects have taken over from band-like behaviour.

ARPES results correspondingly provide due record of the crossover effected between nodally and antinodally dominated gapping of the superconducting state. The data of Mesot *et al* [41], to which often a single *d*-wave cos $2\phi$ fitting to $\Delta(\phi)$ (with or without augmentation by its cos $6\phi$ harmonic) has been pursued, is clearly better viewed as deviating from such formulation around $\phi$ = 30° as apparent in fig.3's centre panel. There are additionally clear complications near and below the hot-spot angle of $\phi$ = 15° as the zone boundary is approached. With Bi-2201 and Na-CCOC the dichotomy between the nodal and antinodal behaviour becomes more evident [42,43] as the two energy scales $\Delta_{sc}$ and $U$ have become more separated – in Bi-2201 due to metallic screening, in Na-CCOC due to excessive ionicity and localization.

Of particular note for the latter system is the relatively low degree of fanning out of the modal dispersion lines in fig.4. Such plots realized via the STS experiment are currently under renewed investigation by Hanaguri and colleagues [44]. The latter have indicated that the relative intensities of the diffraction spots portrayed in figure 2 above respond to a strong magnetic field being applied around the tunnelling tip. Depending upon which zone quadrant the tunnelling electrons relate to, the associated spots are found either to rise or fall in intensity. This raises interesting questions relating to symmetry and to the postulated difference [1b] between extended-*s*, $A_{1g}$ symmetry for the local pairs and $B_{1g}$ $d_{x^2-y^2}$ symmetry for the single quasi-particles and the BCS-like nodal gapping. Details are awaited now with much interest.

## §4. Summary

Close examination has been made of the recent detailed STM investigations from Alldredge *et al* and Hanaguri *et al* on Bi-2212 and Na-CCOC respectively. This examination has been conducted from the perspective of the resonant, boson-fermion, negative-*U*-based crossover mechanism for HTSC, developed over many years by the present author [1,5,11ab,25,*etc*.]. After laying out again the relevant segments of this understanding in §1, a detailed restatement of Alldredge *et al*'s paper has been given in §2, and of Hanaguri *et al*'s linked paper in §3. The first paper deals principally with a determination of gap values and their extreme variability across the field of scan. There clearly are seen to be present two distinct gaps, associated the one with nodal bahaviour (and homogeneous), the other with antinodal physics (and highly inhomogeneous). These observations are readily accounted for in the resonant, two-component, two-stage mechanism proposed. The second paper deals with diffraction features in the optically Fourier transformed gap maps, associated with quasielastic scattering from features probed here at tunnelling voltages that tie this scattering in some way to the superconductivity. Hanaguri *et al*, as have their predecessors, relate this scattering to classical Bogoliubovons. The current paper shows this is untenable and relates the action to scattering by the linearly dispersed, uncondensed local pair mode. How this mode changes with doping level and with system choice has then been explored in detail.

**References**

[1]a  Wilson J A  2004 *Philos. Mag.* **84** 2183.
   b  Wilson J A  2007 *J. Phys.: Condens. Matter* **19** 106224.
   c  Wilson J A  2008 *J. Phys.: Condens. Matter* **20** 015205.
[2]  Pan S H *et al* 2001 *Nature* **413** 282.
   Howald C, Eisaki H, Kaneko N, Greven M and Kapitulnik A  2003 *Phys. Rev.* B **67** 014533.
   Lang K M, Madhavan V, Hoffman J E, Hudson E W, Eisaki H, Uchida S and Davis J C
       2002 *Nature* **415** 412.
[3]  Honma T and Hor P H  *arXiv*:0801.1537.
[4]  Kanigel A, Chatterjee U, Randeria M, Norman M R, Souma S, Shi M, Li Z Z, Raffy H
       and Campuzano J C  2007 *Phys. Rev. Lett.* **99** 157001.
   Yoshida T, Zhou X J, Tanaka K, Yang W L, Hussain Z, Shen Z-X, Fujimori A,
       Sahrakorpi S, Lindroos M, Markiewicz R S, Bansil A, Komiya S, Ando Y, Eisaki H,
         Kakeshita T and Uchida S  2006 *Phys. Rev.* B **74** 224510
   Tanaka K, Lee W S, Lu D H, Fujimori A, Fujii T, Risdiana, Terasaki I, Scalapino D J,
       Devereaux T P, Hussain Z and Shen Z-X  2006 *Science* **314** 1910.
   Lee W S, Vishik I M, Tanaka K, Lu D H, Sasegawa T, Nagaosa N, Devereaux,
       Hussain Z and Shen Z-X,  2008 *arXiv*:0801.2819.
[5]a  Wilson J A and Zahrir A  1997 *Rep. Prog. Phys.* **60** 941.
   b  Wilson J A  2000 *J. Phys.: Condens. Matter* **12** R517.
   c  Wilson J A  2001 *J. Phys.: Condens. Matter* **13** R945.
   d  Wilson J A  2007 *J. Phys.: Condens. Matter* **19** 466210 (10pp).
[6]  Li Y and Lieber C M  1993 *Mod Phys Lett* B **7** 143.
   Li Y, Huang J L and Lieber C M  1992 *Phys Rev Lett* **68** 3240.
[7]  Hayden S M, Mook H A, Dai P, Perring T G and Dogan F  2004 *Nature* **429** 531.
[8]  Wang Y, Ono S, Onose Y, Gu G, Ando Y, Tokura Y, Uchida S and Ong N P
       2003 *Science* **299** 86.
   Li L, Wang Y, Naughton M J, Komiya S, Ono S, Ando Y and Ong N P
       2006 *arXiv:cond-mat*/ 0611731.  (Int. Conf. Mag., Kyoto 2006).
[9]  Friedberg R and Lee T D 1989 *Phys. Rev.* B **40** 6745.
   Domanski T and Ranninger J  2004 *Phys. Rev.* B **70** 184503 and 184513.
   Micnas R, Robaszkiewicz S and Bussmann-Holder A  2005 *Struct. Bond.* **114** 13-69.
   Chen Q, Stajic J and Levin K  2005 *arXiv:cond-mat*/0508603,
       published in  2006 *Low Temp. Phys.* **32** 406. [2006 *Fiz. Nizk.Temp.* **32** 538.]
   de Llano M and Tolmachev V V  2003 *Physica* A **317** 546.
   Garg A, Krishnamurthy H R and Randeria M  2005 *arXiv:cond-mat*/0503144.
   Toschi A, Barone P, Capone M and Castellani C  2005 *New J. Phys.* **7** 7.
[10] Tsuei C C and Kirtley J R  2000 *Rev. Mod. Phys.* **72** 969.
   Tsuei C C, Kirtley J R, Hammerl G, Mannhart J, Raffy H and Li Z Z,

   2004 *Phys. Rev. Lett.* **93** 187004.

[11]a Wilson J A  2006 *J. Phys.: Condens. Matter* **18** R69.

 b Wilson J A  1998 *J. Phys.: Condens. Matter* **10** 3387.

 c Birgeneau R J, Stock C, Tranquada J M and Yamada K

   2006 *J. Phys. Soc. Jpn.* **75** 111003.

[12] Hussey N E  2007 Chap. 10 in 'Handbook on High Temperature Superconductors'

   Ed: Schrieffer J R and Brooks J S ; Pub: Springer Verlag.

 Hussey N E  2008  *J.Phys.:Condens. Matter* **20** 123201.

 Abdel-Jawad M, Kennett M P, Balicas L, Carrington A, McKenzie A P, McKenzie R H

  and Hussey N E  2006 *Nature Phys.* **2** 821.

 Abdel-Jawad M, Analytis J G, Carrington A, French M M J, Hussey N E, Balicas L,

  Charmant J P H  2007 *Phys. Rev. Lett.* **99** 107002

 Hussey N E, Alexander J C and Cooper R A  2006 *Phys. Rev.* B**74** 214508.

[13] Narduzzo A, Albert G, French M M J, Mangkorntong N, Nohara M,Takagi H and

  Hussey N E  *arXiv*:0707.4601

 Ono S, Komiya S and Ando Y  2007 *Phys. Rev.* B  **75** 024515.

 Ando Y  *arXiv*:0711.4213.

[14] Obertelli S D, Cooper J R and Tallon J L  1992 *Phys. Rev. B* **46** 14928.

 Wilson J A and Farbod M  2000  *Supercond. Sci. & Tech.* **13** 307.

[15]  Basov D N and Timusk T  2005 *Rev. Mod. Phys.* **77** 721.

[16]  Casas M, de Llano M, Puente A, Rigo A and Solis M A  2002 *Solid St. Commun.* **123** 101.

[17] Corson J, Mallozo R, Orenstein J, Eckstein J N and Bozovic I  1999 *Nature* **398** 221.

 Corson J, Orenstein J, Oh S, O'Donnell J and Eckstein J N

  2000 *Phys. Rev. Lett.* **85** 2569.

[18] Homes C C, Dordevic S V, Gu G D, Li Q, Valla T and Tranquada J M

  2006 *Phys. Rev. Lett.* **96** 257002.

[19] Peets D C, Mottershead J D F, Wu B, Elfimov I S, Liang R, Hardy W N, Bonn D A,

  Raudsepp M, Ingle N J C and Damascelli A  2007 *New J. Phys.* **9** 28.

[20] Valla T, Federov A V, Lee J, Davis J C and Gu G D  2006 *Science* **314** 1914.

[21] Alldredge J W, Lee J, McElroy K, Wang M, Fujita K, Kohsaka Y, Taylor C, Eisaki H,

  Uchida S, Hirschfeld P J and Davis J C  2008 *Nature Phys.* **4** 319;

   see *arXiv*:0801.0087 for supplementary data.

[22] Hanaguri T, Kohsaka Y, Davis J C, Lupien C, Yamada I, Azuma M, Takano M, Ohishi K,

  Ono M and Takagi H  2007 *Nature Phys.* **3** 865..

[23] McElroy K, Lee J, Slezak J A, Lee D-H, Eisaki H, Uchida S and Davis J C

  2005 *Science* **309** 1048.

[24] Hoffman J E, McElroy K, Lee D-H, Lang K M, Eisaki H, Uchida S and Davis J C,

  2002 *Science* **297**, 1148.

[25] Wilson J A  1988 *J. Phys. C Solid State* **21** 2067.

[26] Krasnov V M, Kovalev A E, Yurgens A and Winkler D,  2001 *Phys. Rev. Lett.* **86** 2657.

Zasadzinski J F, Ozyuzer L, Coffey L, Gray K E, Hinks D G and Kenziora C,
   2006 *Phys. Rev. Lett.* **96** 017004.

[27]  Maggio-Aprile L, Renner C, Erb A, Walker E and Fischer Ø,
   1995 *Phys. Rev. Lett.* **75** 2754.

Kugler M, Fischer Ø, Renner C, Ono S and Ando Y,  2001 *Phys. Rev. Lett.* **86** 4911.

[28]  Loram J W, Mirza K A and Cooper J R,   pp. 77-97  in  *Research Review 1998 HTSC*.
   [Ed: W.Y. Liang; Pub: IRC, Univ. of Cambridge, 1998].

Loram J W, Luo J, Cooper J R, Liang W Y and Tallon J L,
   2000 *Physica* C**341-8**, 831; and 2001 *J. Phys. Chem. Solids*, **62**, 59.

[29]  Oda M, Momono N and Ido M,  2004 *J. Phys. Chem. Sol.* **65** 1381.

[30]  Kanigel A, Norman M R, Randeria M, Chatterjee U, Souma S, Kaminski A, Fretwell H M,
   Rosenkranz S, Shi M, Sato T, Takahashi T, Li Z Z, Raffy H, Kadowaki K, Hinks D,
   Ozyuzer L and Campuzano J C,  2006 *Nature Phys.* **2** 447.

[31]  Shen K M, Ronning F, Lu D H, Baumberger F, Ingle N J C, Lee W S, Meevasana W,
   Kohsaka Y, Azuma M, Takano M, Takagi H and Shen Z-X,  2005 *Science* **307** 901.

[32]  Kondo T, Takeuchi T, Kaminski A, Tsuda S and Shin S,
   2007 *Phys. Rev. Lett.* **98** 267004.

Boyer M C, Wise W D, Chatterjee K, Yi M, Kondo T Takeuchi T, Ikuta H
   and Hudson E W,  2007 *Nature Phys.* **3** 802.

[33]  Clark J W, Khodel V A, Zverev M V and Yakavenko V M,
   2004 *Phys. Repts.* **391**, 123.

Carter E C and Schofield A J,  2004 *Phys. Rev.* B **70**, 045107.

[34]  Miles P A, Kennedy S J, Anderson A R, Gu G D, Russell G J and Koshizuka N,
   1997 *Phys. Rev* B **55** 14632.

[35]  de Llano M and Tolmachev V V,  2003 *Physica* A **317** 546.

Casas M, de Llano M, Puente A, Rigo A and Solis M A,
   2002 *Solid St. Commun* **123** 101.

[36]  Campuzano J C, Ding H, Norman M R, Randeria M, Bellman A F, Tokoya T,
   Takahashi T, Katayama-Yoshida H, Mochiku T and Kadowaki K,
   1996 *Phys. Rev* B **53** R14737.

Kanigel A, Chatterjee U, Randeria M, Norman M R, Koren G, Kadowaki K and
   Campuzano J C,  2008 *arXiv*:0803.3052.

[37]  Tanabe K, Hidaka Y, Karimoto S and Suzuki M,  1996 *Phys. Rev.* B **53** 9348.

[38]  Tornow S, Bulla R, Anders F B and Nitzan A,  *arXiv*:08034066.

[39]  Fruchter L, Raffy H and Li Z Z,  2007 *Phys. Rev.* B **76** 212503.

[40]  Meevasana W, Zhou X J, Sahrakorpi S, Lee W S, Yang W L, Tanaka K, Manella N,
   Yoshida T, Lu D H, Chen Y L, He R H, Lin H, Komiya S, Ando Y, Zhou F, Ti W X,
   Xiong J W, Zhao Z X, Sasegawa T, Kakeshita T, Fujita K, Uchida S, Eisaki H,
   Fujimori A, Hussain Z, Markiewicz R S, Bansil A, Nagaosa N, Zaanen J,
   Devereaux T P and Shen Z-X,  2007 *Phys Rev* B **75** 174506.

[41] Mesot J, Norman M R, Ding H, Randeria M, Campuzano J C, Paramekanti A, Fretwell H M, Kaminski A, Takeuchi T, Yokoya T, Sato T, Takahashi T, Mochiku T and Kadowaki K,  1999 *Phys. Rev. Lett.* **83** 840; 1999 *J. Low Temp. Phys.* **117** 365.

[42] Okada Y, Takeuchi T, Shimoyamada A, Shin S and Ikuta H,  2007 *arXiv:c-m*/0709.0220.

[43] Shen K M, Ronning F, Lu D H, Baumberger F, Ingle N J C, Lee W S, Meevasana W, Kohsaka Y, Azuma M, Takano M, Takagi H and Shen Z-X  2005 *Science* **307** 901:
 - see also for LSCO - Hashimoto M, Yoshida T, Tanaka K, Fujimori A, Okusawa M, Wakimoto S, Yamada K, Kakeshita T, Eisaki H and Uchida S,  2006 *arXiv:c-m*/0610758

[44] Hanaguri T, Kohsaka Y Ono M, Maltseva M, Coleman P, Yamada I, Azuma M, Takano M, Ohishi K and Takagi H,  2008 *Bull. Amer. Phys. Soc.* **53**(2) A10.00010.

**Figure captions**

**Figure 1.** Illustration of hot-point scattering of quasi-particles from positions 5 and 6 on the FS near where the latter intersects the oblique zone-quadrant diagonal out into paired states near the corner of the BZ of maximally anti-bonding (shell filling) character [see ref.5b, fig 3].

**Figure 2a.** The elastic scattering vectors from a suitable point on the Fermi surface to the seven other symmetrically equivalent *k*-space locations. For convenience of drawing the starting vector $k_F'$ falls here at a 'hot-spot'. The siting of $k_F'$ alters as *V* is altered.

**Figure 2b.** The distribution of the diffuse diffraction spots occurring in the optically-derived FT of the tunnelling conductance map g(**r**,*V*) for the case illustrated in fig.2a (where starting vector $k_F'$ falls at a hot spot location). The 28 spots result from 28 stationary waves produced by quasi-elastic scattering between the states indicated in fig.2a. The eight corresponding circuits are identified for the spots issuing from the eight symmetry equivalent starting locations for $k_F'$. The pattern changes of course with tunneling voltage *V*, as the tunnelling electrons seek out the appropriate bound state to enter – identified here as a uncondensed local-pair boson mode state rather than a classical Bogoliubovon state of a gapped BCS-like superconductor.

**Figure 3.** The scanning tunnelling spectroscopy (STS) results (for nearly optimally doped material) from McElroy *et al* on Bi-2212 (<*p*> ≈ 0.15) [23] and from Hanaguri *et al* on Na-CCOC (<*p*> ≈ 0.13) [22] here are displayed on a common scale, in conjunction with projected results for Bi-2201, for which such experiments are yet to be conducted. Through this sequence of materials $T_c$ first rises and then falls under the increased covalency. In each panel a *d*-wave BCS gap plot is included corresponding to the simple behaviour 5½ $kT_c = 2\Delta_{sc}^{max}$ ($\phi = 0, T = 0$). In the case of Bi-2212 the figure displays in addition the 'gapping' $\Delta(\phi)$ as extracted by Mesot *et al* from their ARPES data [40]. Note the clear transfer from nodal to antinodal type behaviour apparent around $\phi = 30°$. The STS data are visible only over the darkened segments of the linear plots immediately below $E_F$. The third panel has been constructed from the first two by means of the various linear extrapolations indicated.

**Figure 4**. Schematic of the evolution in disposition of the local-pair mode with (a) hole doping level in a given HTSC system, and (b) change in covalency and screening between HTSC systems. The fanning of the dispersion lines is tighter and the lines extrapolate towards higher binding energies (*U*) in Na-CCOC, which is the most ionic of the three systems portrayed. In Bi-2201 it would seem that by *p* = 0.28 local pairs will never be stable. When *p* ≈ 0 (*ie.* no metallic screening) all three panels indicate, for the very few pairs then formed, a common binding energy ≈ 85 meV. Gaps greater than this relate to other physics such as checker-boarding, striping and Mott insulation.

X

M

Ph.s.

f1

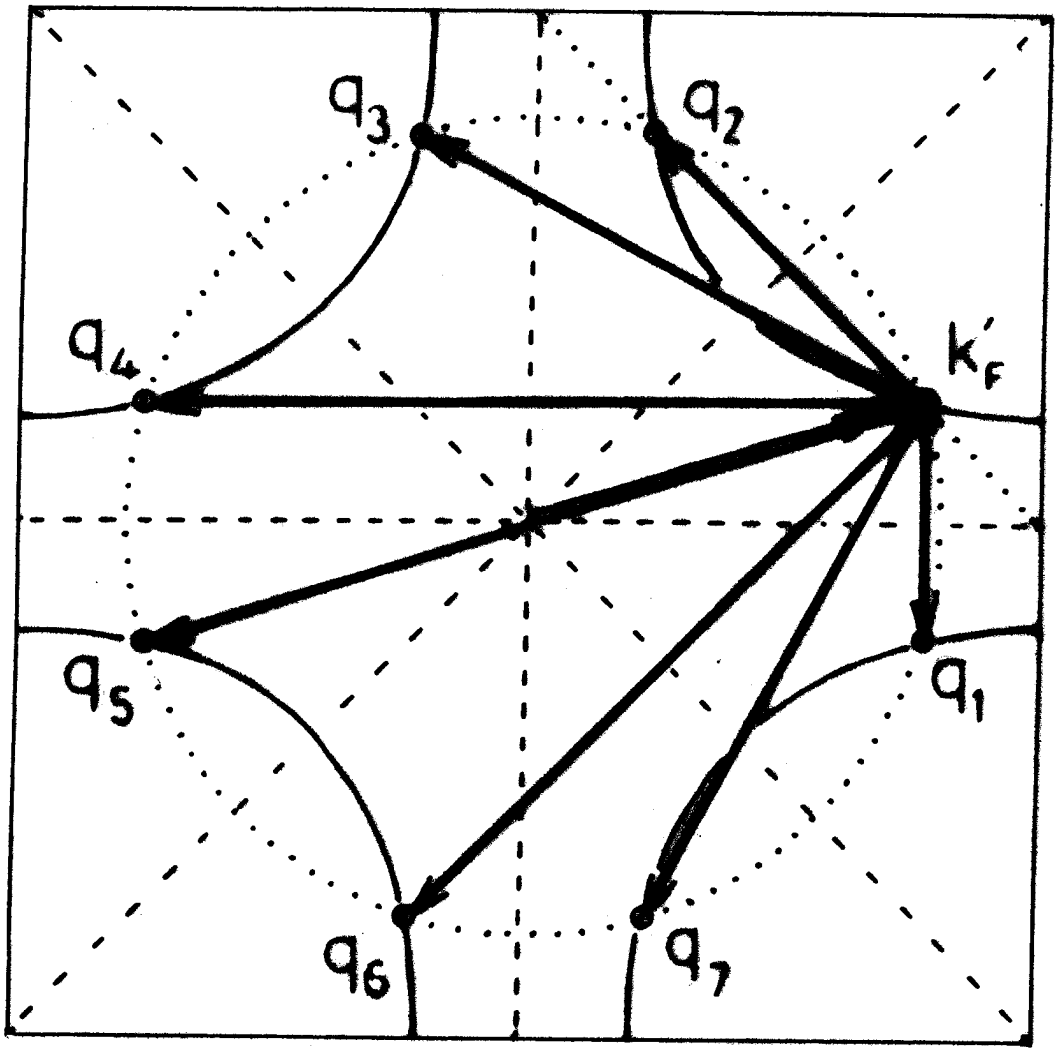

f2a

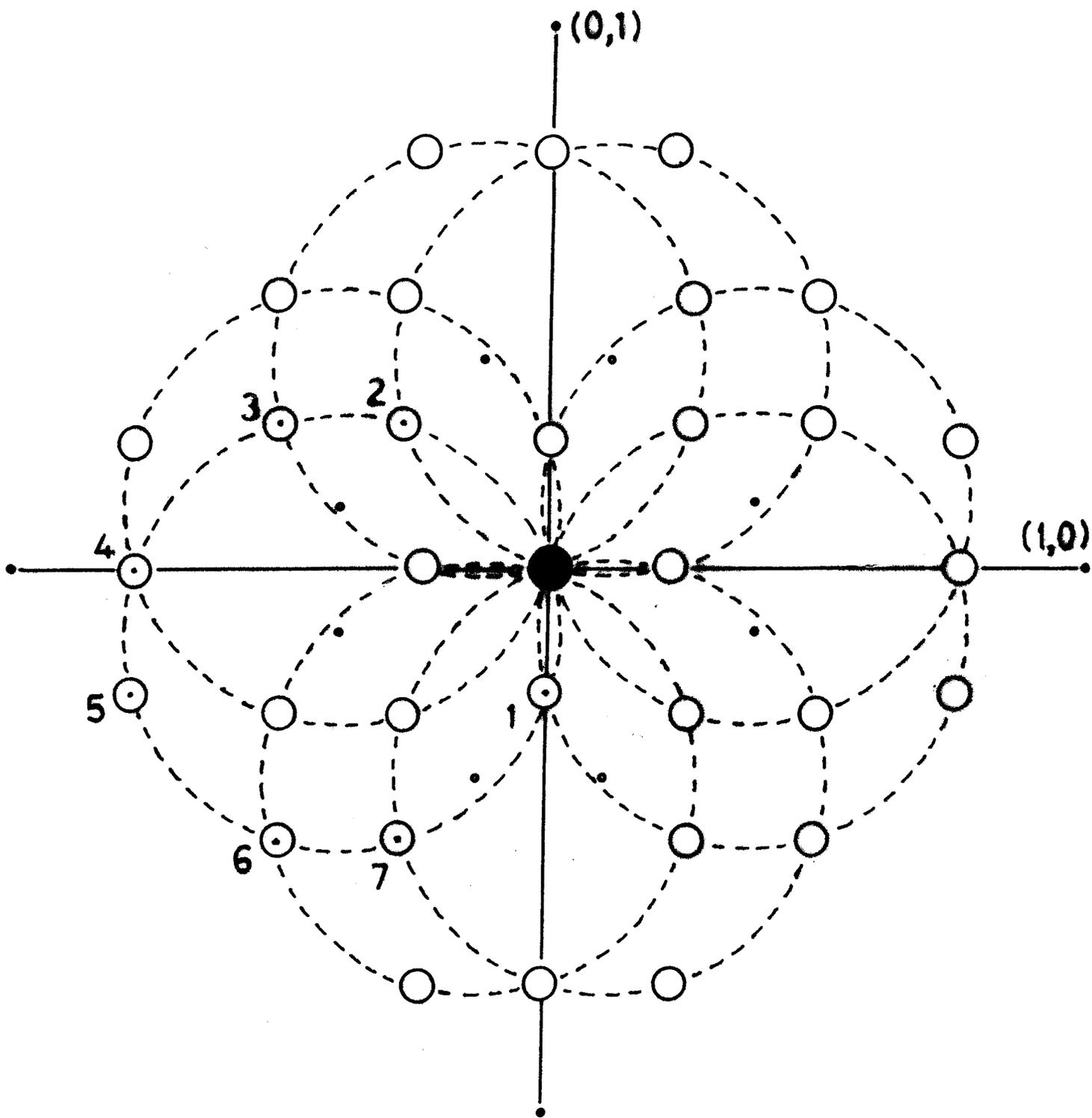

f2b

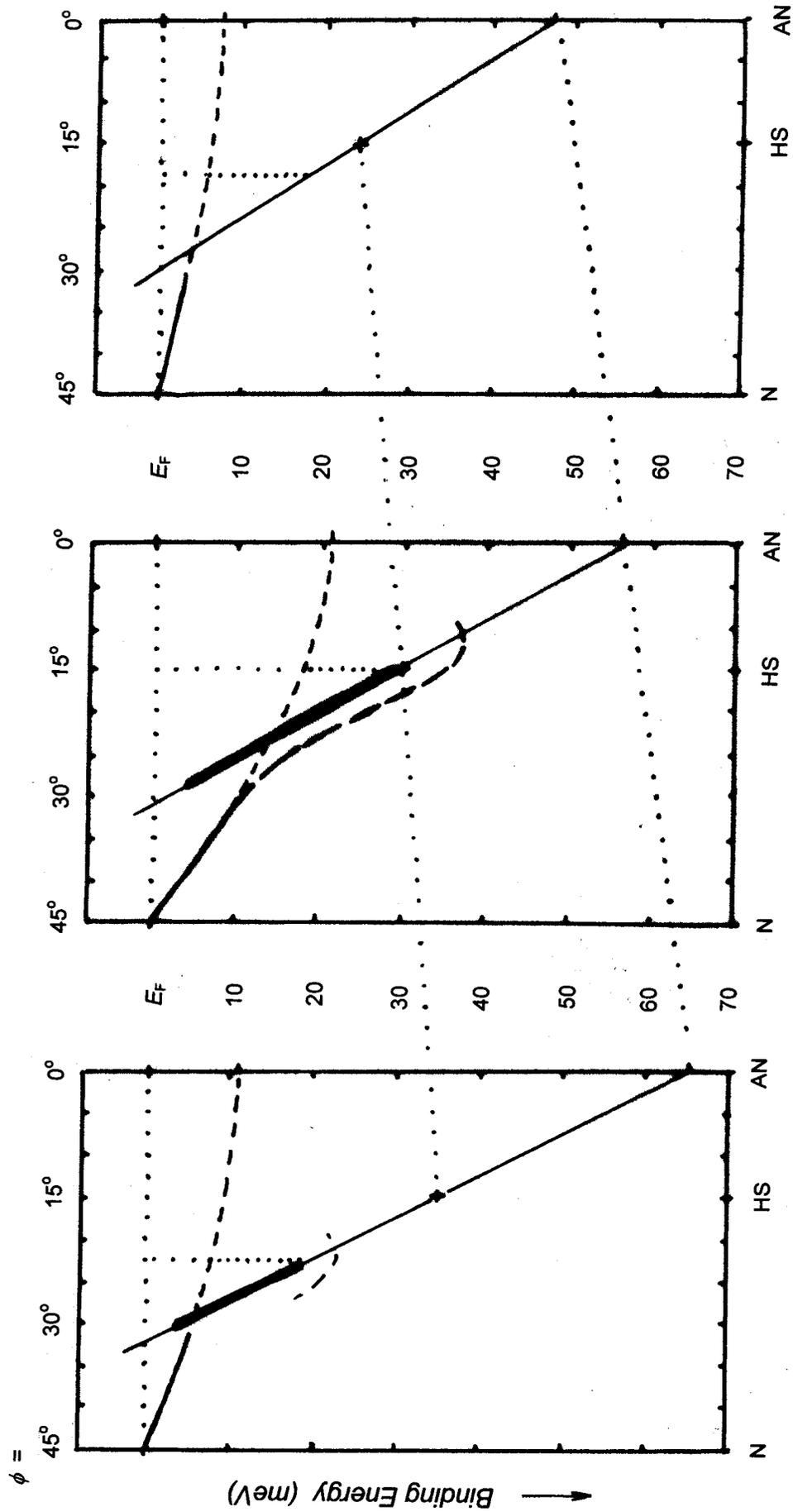

f3

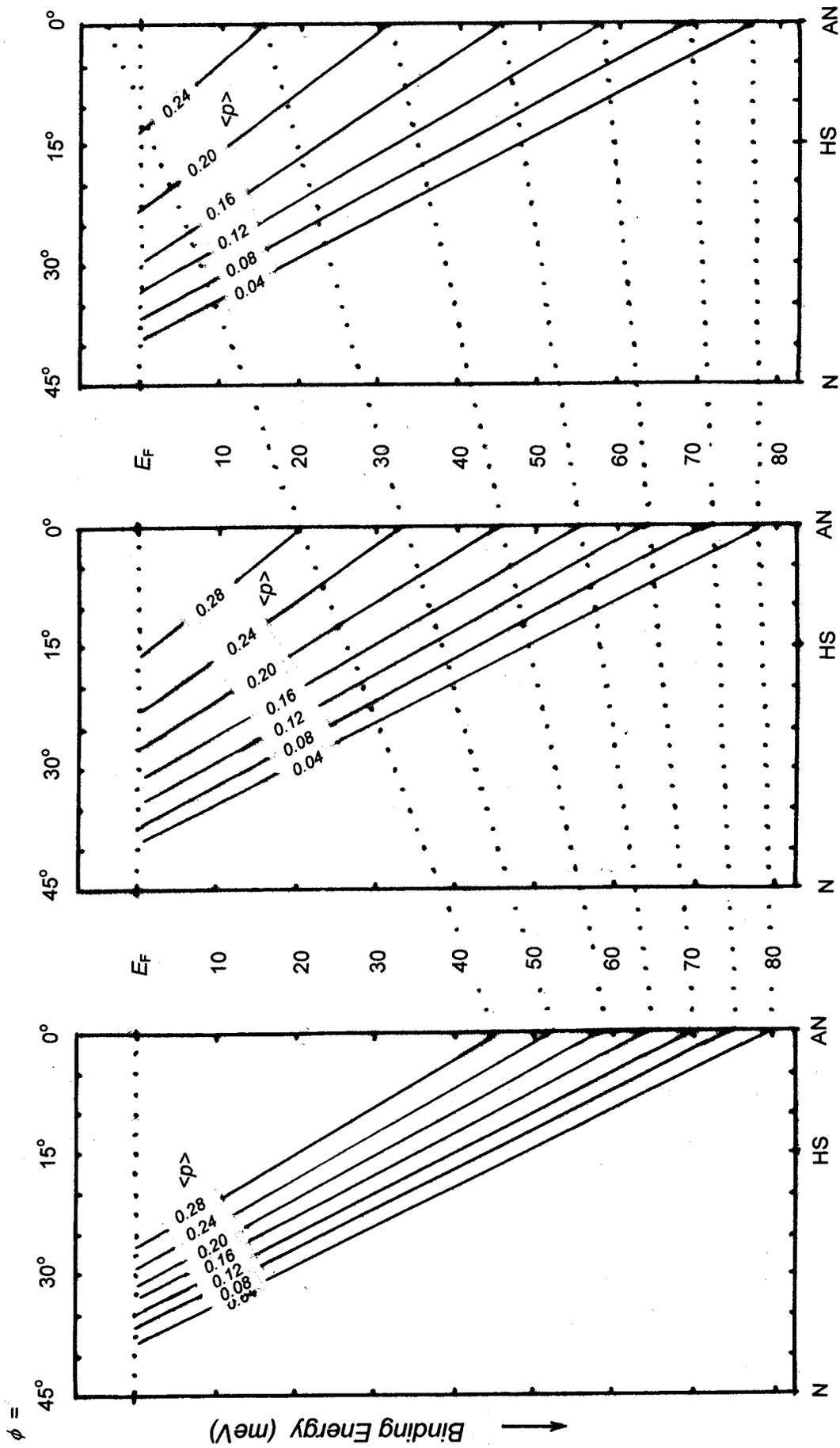

f4